\documentclass{webofc}

\usepackage[utf8]{inputenc} 
\usepackage[T1]{fontenc}    
\usepackage{rotating}
\usepackage{array}
\usepackage{amsmath}
\usepackage[normalem]{ulem}
\usepackage{slashed}
\usepackage{booktabs}
\usepackage[pdftex,table]{xcolor}
\usepackage{units}
\usepackage{xfrac}
\usepackage{mathtools}
\usepackage{empheq}
\usepackage[]{units}
\usepackage{multirow}
\usepackage{amssymb}
\usepackage{url}
\usepackage{comment}
\usepackage{paralist}
\usepackage{xspace}
\usepackage{hyperref}

\usepackage{Commands}

\usepackage{tikz}

\def\beq{\begin{equation}}
\def\eeq{\end{equation}}
\newcommand{\bea}{\begin{eqnarray}\begin{aligned}}
\newcommand{\eea}{\end{aligned}\end{eqnarray}}
\def\bitem{\begin{itemize}}
\def\eitem{\end{itemize}}

\newcommand{\convtune}{\texttt{MC-Conv-Tune}\xspace}
\newcommand{\noerrtune}{\texttt{MC-Tune-NoError}\xspace}
\newcommand{\errtune}{\texttt{MC-Tune-Error}\xspace}

\begin{document}

\title{Event Generator Tuning Incorporating Systematic Uncertainty}

\author{
\firstname{Jaffae} \lastname{Schroff}\inst{1}\fnsep\thanks{\email{jeffae@berkeley.edu}} 
\and
\firstname{Xiangyang} \lastname{Ju}\inst{2}\fnsep\thanks{\email{xju@lbl.gov}} 
}

\institute{
Physics Division, University of California, Berkeley, CA 94720
\and
Scientific Data Division, Lawrence Berkeley National Laboratory, Berkeley, CA 94720
}

\abstract{
Event generators play an important role in all physics programs at the Large Hadron Collider and beyond. Dedicated efforts are required to tune the parameters of event generators to accurately describe data. There are many tuning methods ranging from expert-based manual tuning to surrogate function-based semi-automatic tuning, to machine learning-based re-weighting. Although they scale differently with the number of generator parameters and the number of experimental observables, these methods are effective in finding optimal generator parameters. However, none of these tuning methods includes the Monte Carlo (MC) systematic uncertainties. That makes the tuning results sensitive to systematic variations. In this work, we introduce a novel method to incorporate the MC systematic uncertainties into the tuning procedure and to quantitatively evaluate uncertainties associated with the tuned parameters. Tested with a dummy example, the method results in a $\chi^2$ distribution that is centered around one, the optimal generator parameters are closer to the true parameters, and the estimated uncertainties are more accurate.
}

\maketitle

\section{Introduction}
General-purpose event generators, like Pythia 8, are widely used in High Energy Physics for event generation and physics simulations. They often contain many parameters that must be tuned so that the generated distributions match the data. Dedicated tuning campaigns were launched by the ATLAS and CMS experiments to tune these event generators for the Large Hadron Collider (LHC). 

The tuning method evolved from manual tuning to automated tuning. In the beginning, the tuning was performed by domain experts based on their sense of physics and goodness of fit~\cite{Skands:2014pea}. Later on, the software, Professor~\cite{Buckley:2009bj}, made the tuning automated and more objective. It first optimizes a surrogate function that models the relationship between generator parameters and experimental variables (inner-loop optimization), and then optimizes a $\chi^2$ function that measures the differences between simulated data and experimental data. Recently, Apprentice~\cite{Krishnamoorthy:2021nwv}, a purely Python-based tool, was developed to
leverage High-Performance Computing and introduced rational approximation as an alternative surrogate function. 

However, the Monte Carlo (MC) systematic uncertainties are either ignored or artificially compensated. Ref.~\cite{Skands:2014pea} artificially introduced a 5\% uncertainty when calculating the $\chi^2$ function for experimental histograms so that the $\chi^2$ is not too large, while Professor and Apprentices ignored MC uncertainties. Because of the absence of MC uncertainties, these tunings are often sensitive to systematic variations. For example, the latest ATLAS tuning~\cite{a14tune} finds that tuning with different parton distribution functions (PDFs) results in different tuned parameters.  

Two major sources of MC systematic uncertainties exist: QCD scale and parton distribution functions (PDF). The QCD scale uncertainties stem from the choice of factorize and renormalization QCD scales, while the PDF uncertainties are from either the PDF sets themselves or the differences among PDF sets.

The developments of the LHE 3 data format automate the estimation of MC systematic uncertainties, thanks to the multiple event weights stored in LHE 3 files. We propose to improve the current MC tuning procedure by taking into account these theoretical uncertainties and estimating the parameter uncertainties based on the $\chi^2$ distribution. 

\section{Current MC tuning procedure}
\label{sec:current-tuning}
The current MC tuning procedure is a two-step optimization process, detailed in Refs~\cite{Buckley:2009bj,Krishnamoorthy:2021nwv}. In the inner loop, a surrogate function is optimized to model the relationship between the generator parameters and the experimental observables. In the outer loop, the generator parameters are optimized to minimize a $\chi^2$ function. We will describe the two steps and refer to it as \noerrtune in the following sections.

\subsection{Inner loop optimization}
To illustrate the method, we assume there are $n$ generator parameters and $\ell$ generator parameters are sampled for simulation. The inner loop optimization is performed for each bin of each experimental observable. Without a loss of generality, we focus on one bin and use the quadratic approximation:
\begin{align}
    f(\vect{p}) &= a_{0}^{(0)} + \sum_{i=1}^{n}\,a_{i}^{(1)} p_i 
                   + \sum_{i=1}^{n}\sum_{j=i}^{n}\,a_{ij}^{(2)} p_i p_j \label{equ:poly_approx1} \\
                &= M(\vect{p}) \label{equ:poly_approx2}
\end{align}
where $M(\vect{p})$ is the vector of model predictions corresponding to the parameters $\vect{p}$ and $a_{i(j)}^{(k)}$ are the coefficients of the surrogate function. Equation~\ref{equ:poly_approx1} can be written in a matrix form:
\begin{equation}
    f(\vect{p}) = \vect{P} \cdot \vect{A} = \vect{M} \label{equ:poly_approx3} 
\end{equation}
where \vect{P} is a matrix in which column $k$ contains the parameter variations of model set $k$:
\[
 P_{k,1\dots m} = (1, p_{1}{(k)}, \dots,  p_{n}^{(k)}, \dots,  p_{1}^{(k)2}, \dots,  p_{n}^{(k)2},  p_{1}^{(k)} p_{2}^{(k)}, \dots, p_{n-1}^{(k)\cdot  p_{n}^{(k)}})
\]
The $m = 1 + n + n(n+1)/2$ coefficients $a^{(0, 1, 2)}$ of the surrogate function are unknown and determined by fitting Eq.~\eqref{equ:poly_approx3} to $\ell$ simulation distributions ($\ell \ge m$), generated with different parameter settings.

Solving equation~\eqref{equ:poly_approx3} is to minimize the loss function:
\[
\loss = ||\vect{M} - \vect{P}\cdot\vect{A}||^2 \label{eq:inner_obj}
\]
It can be solved by inverting the matrix $\vec{P}$. Often is the case that there are more experimental runs than the number of coefficients, making the equation over-determined. Therefore, a simple matrix inversion based on singular value decomposition may not be robust. We find adding a penalty term such as lasso or ridge helps to stabilize the optimization process. 

\subsection{Outer loop optimization}
After the surrogate function is optimized, the next step is to optimize the generator parameters by minimizing the $\chi^2$ function:
\[
\chi^2 = \sum_{i}^{B}\frac{[d_i - f(\vect{p}, x_i)]^2}{\sigma_{d_i}^{2}}
\]
where $i$ loops over all $B$ bins, $f_i(\vect{p})$ is the surrogate function for the $i$-th bin, $d_i$ is the experimental measurement, and $\sigma_{d_i}$ is the uncertainty of the measurement. Throughout the procedure, no MC uncertainties are taken into account.

\section{Tuning with MC uncertainties}
A straightforward to incorporate the MC uncertainties is to use another surrogate function $g(\vect{\vect{p}})$ to model the relationship between the generator parameters and the MC uncertainties $e_\text{MC}$ for each bin. This surrogate function can be obtained similarly to the one for the nominal values. Then, the $\chi^2$ function can be modified to include the MC uncertainties:
\[
\chi^2 = \sum_{i}^{B}\frac{[d_i - f(\vect{p}, x_i)]^2}{\sigma_{d_i}^{2} + g(\vect{p}, x_i)^2}
\]
This method is referred to as \errtune in the following sections.

We propose propagating the MC uncertainties to the surrogate function and incorporating that in the $\chi^2$ function. We minimize the following loss function and obtain the covariance matrix $\Sigma$:
\[
    \loss = ||\vect{M} - \vect{P}\cdot\vect{A}||^2 / \vect{M}_{\text{error}}^{2} \label{eq:inner_obj2}
\]
where $\vect{M}_{\text{error}}$ is the MC uncertainties associated with the event generators. With the inner optimization, we obtain not only the coefficients $\vect{A}$ but also the covariance matrix $\Sigma$. The covariance matrix $\Sigma$ is then used to estimate the surrogate function uncertainties $\sigma_{f_i}$ in the $\chi^2$ function:
\[
\sigma_{f}^{2}(\vect{p}) = \vect{J}\Sigma\vect{J}^T
\]
where \vect{J} is the Jacobian matrix of the surrogate function. Now, we can modify the $\chi^2$ function to take into account the MC uncertainties:
\[
\chi^2 = \sum_{i}^{n}\frac{[d_i - f(\vect{p}, x_i)]^2}{\sigma_{d_i}^{2} + \sigma_{f_i}^2(\vect{p})}
\]
This method is referred to as \convtune in the following sections.

\section{Toy data setup}
\begin{figure}
    \centering
    \includegraphics[width=0.8\textwidth]{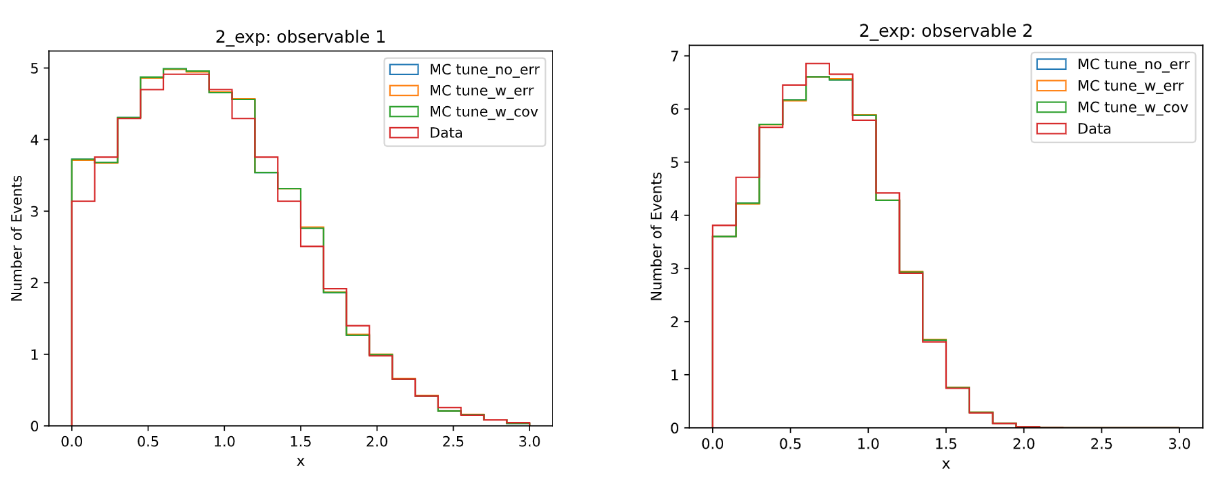}
    \caption{Toy observables. The red curve labeled as "Data" is the target distribution. Other curves, labeled as "MC" are the distributions generated by generator parameters tuned with different methods as detailed in the text.
    }
    \label{fig:observables}
\end{figure}

We create toy data to evaluate the effectiveness of our method. We define two observables following exponential functions: 
\[
    y_0 = e^{a x_0 + b x_{0}^{2}}, \quad
    y_1 = e^{a x_1 + b x_{1}^{3}}
\]
$a$ and $b$ are two generator parameters that control the observable distributions.
Like in practices, generator parameters are often bounded by physical constraints, we set the boundaries of $a$ to be $[1, 2]$ and $b$ to be $[-1.2, 0.8]$. Like the experimental measurements, these toy observables are histograms with 20 bins, as shown in a red curve in Fig.~\ref{fig:observables}.

Following the tuning procedure outlined in Section~\ref{sec:current-tuning}, we randomly sample 30 independent pairs of $(a, b)$ with 100,000 events for each pair. We then use the 3rd-order polynomial function as the surrogate function for all three tuning methods.

\section{Results}
\begin{figure}
    \centering
    \includegraphics[width=0.8\textwidth]{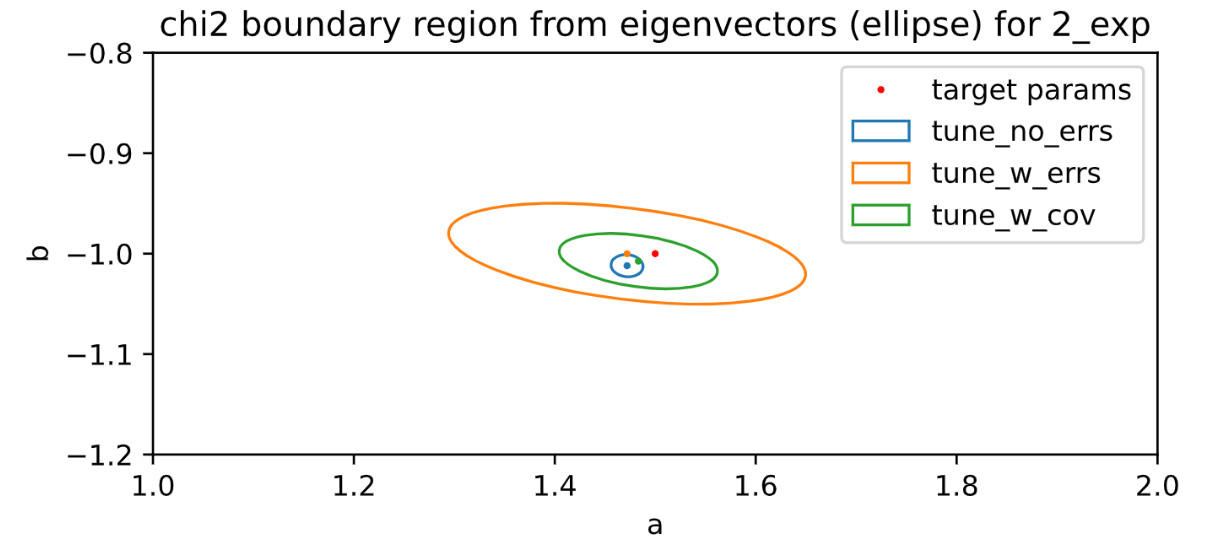}
    \caption{68\% confidence level contour in the $a$ and $b$ plane. The solid dots are the optimal generator parameters obtained with different methods.}
    \label{fig:objective-2d}
\end{figure}

Figure~\ref{fig:observables} compares the toy observables between the target distributions and the tuned ones. We see that all three methods can obtain optimal generator parameters that produce distributions that agree with the target distribution. The optimal and true generate parameters are displayed in Fig~\ref{fig:objective-2d}. The \convtune finds the optimal parameters closest to the true parameters because it takes into account the MC uncertainties properly. In addition, we draw a contour where the objective function is larger than their minimum values by the number of degrees of freedom. The \noerrtune yields a very narrow contour, indicating the estimated errors are underestimated. That confirms the findings from Ref~\cite{Buckley:2009bj}, where the authors did not use the number of degrees of freedom to estimate the errors but instead used an educated guess of the threshold [see the Eigentune method]. On the other hand, the \errtune yields a very wide contour, indicating the estimated errors are overestimated. This is because the observable values and their errors should not be modeled with independent surrogate functions. The \convtune yields a contour that is in between the other two methods and encompasses the true parameters.

\begin{figure}
    \centering
    \includegraphics[width=0.6\textwidth]{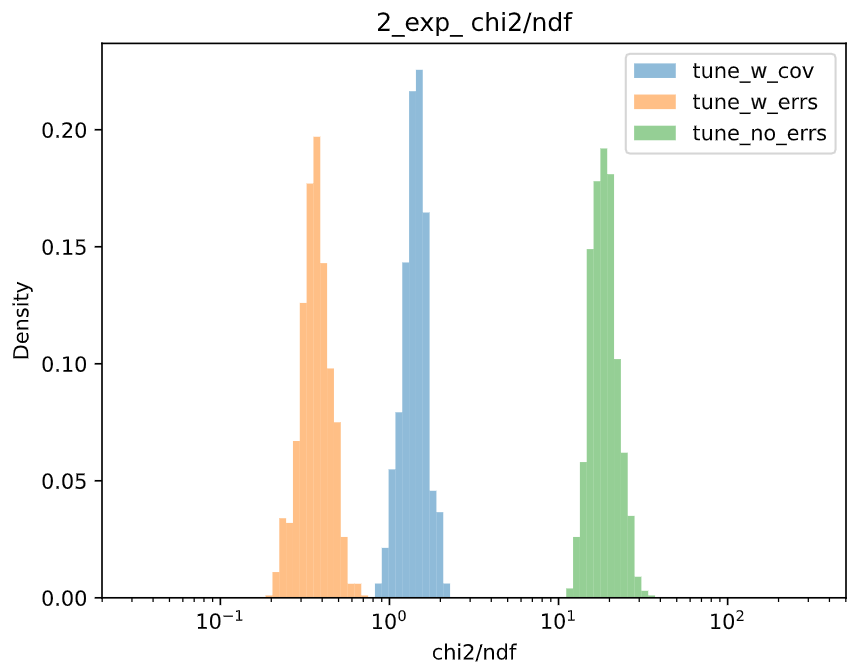}
    \caption{$\chi^2$ over the number of degrees of freedom for different tuning methods obtained with 100 trials.}
    \label{fig:objective-1d}
\end{figure}

To check the stability of the tuning methods, we repeat the tuning procedure 100 times. Figure~\ref{fig:objective-1d} shows the $\chi^2$ distribution over the number of degrees of freedom. All algorithms yield a relatively narrow width. However, the \errtune and \noerrtune have a slightly larger tail fraction. As inferred from Fig.~\ref{fig:objective-2d}, the \convtune peaks around one, while the other two methods yield either much larger or smaller values. 

\section{Conclusion}
We propose a new method to incorporate the MC systematic uncertainties into the MC tuning procedure. We evaluate the method with a toy example and find that the method yields better optimal generator parameters and uncertainty estimations. The method can be easily extended to include different sources of uncertainties. 

MC uncertainties are often independent of the experimental uncertainties. Thanks to recent developments of the HepData repo and the support of LHC experiments, the LHC experiments started to report the breakdown of their measurement uncertainties into theoretical and experimental uncertainties. Within our method, we can properly correlate the MC uncertainties with the reported theoretical uncertainties, and uncorrelate them with the experimental uncertainties. Doing so will further improve the error estimations.

However, this method is computationally expensive. It is much slower than the current MC tuning procedure. We are working on parallelizing the optimization process with GPUs or multithreading in CPUs.

\bibliography{References}

\begin{thebibliography}{4}

\bibitem{Skands:2014pea}
P.~Skands, S.~Carrazza, J.~Rojo, Eur. Phys. J. C \textbf{74}, 3024 (2014),
  \texttt{1404.5630}

\bibitem{Buckley:2009bj}
A.~Buckley, H.~Hoeth, H.~Lacker, H.~Schulz, J.E. von Seggern, Eur. Phys. J.
  \textbf{C65}, 331 (2010), \texttt{0907.2973}

\bibitem{Krishnamoorthy:2021nwv}
M.~Krishnamoorthy, H.~Schulz, X.~Ju, W.~Wang, S.~Leyffer, Z.~Marshall,
  S.~Mrenna, J.~M\"uller, J.B. Kowalkowski, EPJ Web Conf. \textbf{251}, 03060
  (2021), \texttt{2103.05748}

\bibitem{a14tune}
{ATLAS Collaboration}, \emph{{ATLAS Pythia 8 tunes to 7 TeV data}},
  {ATL-PHYS-PUB-2014-021} (2014)

\end{thebibliography}
\end{document}